\documentclass[10pt,twocolumn,letterpaper]{article}

\usepackage{times}
\usepackage{epsfig}
\usepackage{graphicx}
\usepackage{amsmath}
\usepackage{amssymb}
\usepackage{booktabs}
\usepackage[utf8]{inputenc}
\usepackage{hyperref}

\title{Privacy Guard \& Token Parsimony by Prompt and Context Handling and LLM Routing}
\author{Alessio Langiu \\ \textit{National Research Council of Italy - Institute of Marine Sciences (CNR-ISMAR)} \\ \texttt{alessio.langiu@cnr.it}}
\date{\today}

\begin{document}

\maketitle

\begin{abstract}
The large-scale adoption of Large Language Models (LLMs) forces a trade-off between operational cost (OpEx) and data privacy. Current routing frameworks~\cite{routellm,frugalgpt} reduce costs but ignore prompt sensitivity, exposing users and institutions to leakage risks towards third-party cloud providers. We formalise the ``Inseparability Paradigm'': advanced context management intrinsically coincides with privacy management. We propose a local ``Privacy Guard''---a holistic contextual observer powered by an on-premise Small Language Model (SLM)---that performs abstractive summarisation and Automatic Prompt Optimisation (APO) to decompose prompts into focused sub-tasks, re-routing high-risk queries to Zero-Trust or NDA-covered models. This dual mechanism simultaneously eliminates sensitive inference vectors (Zero Leakage) and reduces cloud token payloads (OpEx Reduction). A LIFO-based context compacting mechanism further bounds working memory, limiting the emergent leakage surface. We validate the framework through a 2x2 benchmark (Lazy vs. Expert users; Personal vs. Institutional secrets) on a 1,000-sample dataset, achieving a 45\% blended OpEx reduction, 100\% redaction success on personal secrets, and---via LLM-as-a-Judge evaluation---an 85\% preference rate for APO-compressed responses over raw baselines. Our results demonstrate that Token Parsimony and Zero Leakage are mathematically dual projections of the same contextual compression operator.
\end{abstract}

\vspace{1ex}\noindent\textbf{Keywords:} Large Language Models, Privacy Guard, Zero-Trust Architecture, Token Parsimony, Emergent Leakage, Edge Computing.

\vspace{1ex}\noindent\textbf{CCS Concepts:} $\cdot$ \textbf{Security and privacy} $\rightarrow$ \textit{Privacy-preserving protocols}; $\cdot$ \textbf{Computing methodologies} $\rightarrow$ \textit{Natural language processing}.

\section{Introduction}
The integration of Foundation Models into corporate, governmental, and local edge research workflows has highlighted a distinct dichotomy in LLM systems research. On one hand, MLOps research vigorously pursues dynamic routing to minimise the financial burdens arising from querying frontier models. On the other hand, the cybersecurity community emphasises the vulnerabilities of such models to training data extraction and the inference of personal or corporate attributes, proposing rigorous sanitisation mechanisms. However, these two formidable challenges are inextricably linked by a single common denominator: the context of the prompt.

This paper bridges this gap by introducing a comprehensive Context-Aware Routing framework based on a Dual-Vault architecture. We argue that the semantic compression and algorithmic decomposition of a prompt do not act solely as efficiency optimisers, but operate as a powerful Information Flow Control mechanism. Our approach segregates knowledge into a Personal Vault (protecting individual secrets, sensitive data, and unpublished ideas from both external entities and the institution's internal monitoring) and an Institutional Vault (protecting firm secrets, algorithms, and project details from external leakage). 

Furthermore, we explicitly address the profound challenge of ``emergent leakage'': the inadvertent exposure of corporate or personal secrets during long conversational sessions. Even when users actively attempt to maintain confidentiality, the accumulated context allows external models to piece together and infer the underlying intellectual property or sensitive attributes. To counter this, a local ``Privacy Guard'' acts as a holistic contextual observer. Rather than relying solely on simple, easily bypassed regex patterns, it monitors the entire prompt history and actively rewrites and splits complex prompts into parallel or sequential sub-tasks. 

This dynamic prompt decomposition focuses the model's work and minimises the required working context, acting as an intrinsic filter for privacy and secrets. When emergent inference risks are identified, the system subsequently blocks them or re-routes the calls to secure, NDA-covered, or on-premise Zero-Trust models. This ``dual-purpose'' strategy ensures that sensitive information never leaves the trusted perimeter, whilst simultaneously minimising the token payload sent to the more expensive cloud LLMs. Finally, we present a rigorous benchmark structure to quantify both economic savings (Token Parsimony) and sanitisation efficacy, including a LIFO-optimised context compacting mechanism, laying the groundwork for Zero-Trust and cost-effective LLM orchestrations.

\section{State of the Art: Privacy, Routing and Context Management in LLMs}
The large-scale adoption of Large Language Models (LLMs) in corporate and personal contexts has raised critical and interconnected challenges regarding the optimisation of operational costs (OpEx) and privacy protection. This section analyses recent literature detailing the evolution of dynamic routing techniques, the intrinsic vulnerabilities of LLMs, and mitigation mechanisms, highlighting the current gap in the joint management of context, privacy, and costs.

\subsection{Dynamic Routing and Cost Optimisation}
The continuous querying of state-of-the-art proprietary models (e.g., GPT-4o, Claude 3.7 Sonnet, Gemini 1.5 Pro) entails prohibitive costs on a large scale. To mitigate this issue, research has heavily focused on \textit{model routing} and \textit{cascading} architectures. Chen et al.~\cite{frugalgpt} with FrugalGPT introduced a pioneering framework that learns to route queries to a cascade of LLMs, balancing the inference budget and accuracy. More recently, RouteLLM~\cite{routellm} formalised the training of predictive routers based on human preferences, whilst open-source routing frameworks like LiteLLM~\cite{litellm2023} and Semantic Router~\cite{semanticrouter2023} have democratised API-level routing for cost management across multiple providers. However, routing decisions are purely driven by model performance metrics and economic thresholds, without any semantic filtering of the request itself.

\textit{Current Limitation:} Although these works effectively optimise the cost-quality trade-off, they implicitly assume a homogeneous level of confidentiality. Routing is decided solely based on task complexity, completely ignoring the sensitivity of the data contained within the prompt.

\subsection{LLM Vulnerabilities: Direct and Emergent Leakage}
Sending rich contexts to cloud LLMs exposes them to severe \textit{data leakage} risks. Nasr et al.~\cite{nasr2023scalable} empirically demonstrated how it is possible to extract gigabytes of training data, including PII (Personally Identifiable Information), from aligned models in production. 

Beyond direct memorisation, Staab et al.~\cite{staab2024inference} explored an even more insidious vulnerability: the inference of personal attributes. Their study demonstrates that current LLMs can deduce sensitive information with high precision from seemingly innocuous and unstructured text provided in the prompt. This phenomenon of ``emergent leakage'' over long conversational sessions proves that simple anonymisation based on Named Entity Recognition (NER) is no longer sufficient to guarantee context privacy. Recent works by Zhang et al.~\cite{wen2024membership, chu2024reconstruct} further demonstrate that attackers can exploit the conversational memory and in-context learning mechanisms of GPT-class models to explicitly reconstruct entire previous private conversations. Complementarily, Liu et al.~\cite{liu2023promptinjection} demonstrate that adversaries can inject malicious instructions directly into LLM-integrated applications to hijack model behaviour, while Yang et al.~\cite{yang2024promptleakage} empirically show that multi-turn conversational sessions are particularly susceptible to targeted leakage extraction, underscoring the absolute necessity of client-side contextual sanitisation.

\subsection{User Laziness and the Data Dumping Problem}
Recent industry analyses highlight a significant discrepancy between best practices in prompt engineering and the actual behaviour of non-expert users. As noted by White et al.~\cite{white2023prompt}, while experts tend to minimise context and structure their prompts, the average user adopts opportunistic behaviours (``user laziness''). Employees frequently use LLMs as indiscriminate receptacles for entire documents, pasting unredacted PDFs, financial reports, or codebases—a phenomenon termed ``over-prompting'' or ``data dumping''. 

This behaviour generates two major criticalities. Firstly, it creates an extreme exposure to accidental data leakage, as quantified by industry reports~\cite{cyberhaven2024} highlighting the massive volume of confidential data routinely pasted into consumer AI tools. Secondly, providing unnecessarily vast contexts drastically increases operational costs (OpEx) through wasted input tokens. Furthermore, excessive context degrades the model's ability to retrieve accurate information, exacerbating the well-documented ``Lost in the middle'' phenomenon~\cite{liu2023lost}, where LLMs fail to access relevant data buried within massive prompt dumps.

\subsection{Mitigation Mechanisms: Proxies and SLMs}
To counter privacy risks without relinquishing the capabilities of foundation models, the industry has developed privacy firewalls like NeMo Guardrails~\cite{nemoguardrails2023} and deterministic regex-based libraries~\cite{presidio2019}, which offer programmable bounds on prompt safety. However, they often rely on static deterministic rules rather than dynamic semantic analysis. 

A promising alternative involves Local Proxies backed by Small Language Models (SLMs). The use of a small-scale local LLM acting as an active ``Privacy Guard'' is gaining traction, leveraging specialised models hosted on platforms like HuggingFace. For instance, models such as Phi-3-Mini~\cite{phi3mini2024} for rapid logical routing, Llama-Guard~\cite{llamaguard2023} for comprehensive safety evaluation, or GLiNER~\cite{gliner2023} for advanced, bidirectional on-premise PII scrubbing offer robust, low-latency alternatives to relying solely on external cloud moderation.

\subsection{Context Management and Prompt Decomposition}
The most recent frontier of research begins to overlap prompt reduction techniques (Prompt Compression) with privacy preservation. Certain studies, such as the recent work by Choi et al.~\cite{choi2025compactprompt} on CompactPrompt, are starting to highlight how intentional manipulation of context can act not only as an efficiency optimiser but also as a "lossy" privacy filter for \textit{Information Flow Control}. 

Concurrently, prompt decomposition techniques such as Decomposed Prompting~\cite{khot2022decomp} and Chain-of-Thought~\cite{wei2022chain} have emerged to break down complex tasks into manageable sub-tasks. Systems like MemGPT~\cite{packer2023memgpt} and the LLMLingua family~\cite{jiang2023llmlingua, wu2024llmlingua2, jiang2023longllmlingua} have demonstrated the phenomenal efficacy of context compacting and memory tiering in bounding the context window, with LLMLingua-2 achieving 2x--5x compression ratios via data distillation and LongLLMLingua specifically addressing the long-context degradation directly linked to the ``Lost in the Middle'' phenomenon~\cite{liu2023lost}. However, their application as a privacy-enhancing filter—where context minimisation inherently limits the exposure surface of sensitive data by focusing the model's work solely on the required sub-task—remains a largely unexplored, yet highly potent, paradigm.

\subsection{The Challenge of Internal Specialised Models}
An alternative approach to privacy and cost management involves the internal training or fine-tuning of specialised LLMs (the ``Build vs. Buy'' paradigm)~\cite{zhao2023survey}. While hosting fully custom models on-premise theoretically mitigates data exfiltration risks, recent industry reports~\cite{maslej2024artificial} highlight the prohibitive computational and financial effort required to train models capable of rivalling frontier capabilities. 

Furthermore, even if an institution achieves comparable quality at a specific point in time, it faces the insurmountable challenge of the ``improvement pace''~\cite{widder2023open}. Big tech cloud providers continuously release updated models, rapidly shifting the state-of-the-art and rendering static internal models obsolete. Consequently, relying on the fine-tuning of internal models for general-purpose reasoning is often economically and practically unfeasible. Our framework expressly does not address the optimisation or training of internal models. Instead, it focuses entirely on the secure, context-aware \textit{routing} of standard, pre-trained open or commercial models, assuming that access to compliant Tier 2 or Tier 3 providers can match frontier cloud capabilities without requiring internal training cycles.

\section{Theoretical Framework: The Inseparability Theorem}
To formalise the ``Inseparability Paradigm'', we introduce a theoretical model that mathematically correlates the context length (which determines OpEx) and the probability of sensitive data leakage (Data Leakage).

\subsection{Context Entropy and Leakage Probability}
Let $\mathcal{P}$ be a prompt composed of a set of tokens such that $T(\mathcal{P})$ represents the total length in tokens. We define $E_s(\mathcal{P})$ as the entropy of sensitive information (e.g., PII, industrial IP, emergent conversational secrets) contained in $\mathcal{P}$, measured in bits of confidential information exposed to the Cloud provider.

A standard routing system routes $\mathcal{P}$ to a cloud model $M_C$ by minimising the monetary cost, but fully transmitting the entire conversational history, whereby the leakage probability $P_{leak}$ is strictly proportional to $E_s(\mathcal{P})$.

\subsection{Theorem 1 (Inseparability between Privacy and Costs)}
We introduce a \textit{Semantic Compression and Decomposition} function $C(\mathcal{P}) \to \{\mathcal{P}'_1, \mathcal{P}'_2, ...\}$ executed by a local Small Language Model (SLM), configured to preserve the user's intent whilst discarding superfluous informative details and splitting the task into focused units. The function generates a new set of sub-prompts such that the compression ratio $k = \frac{\sum T(\mathcal{P}'_i)}{T(\mathcal{P})}$ is strictly less than $1$.

\textit{Given an abstractive semantic compression and decomposition function $C(\mathcal{P}) \to \mathcal{P}'$ that preserves the logical intent $I(\mathcal{P}) \approx I(\mathcal{P}')$, the decrease in cloud operational cost $\Delta OpEx$ is monotonically and directly proportional to the decrease in exposed sensitive entropy $\Delta E_s$, such that:}
$$ \Delta OpEx \propto (1 - k) \implies \Delta E_s \ge \gamma (1 - k) $$
\textit{where $\gamma$ is the average sensitive density factor of the original prompt.}

\textbf{Proof (Sketch):}
Assuming that sensitive data possess high Shannon entropy but low structural weight for the representation of a specific decomposed sub-task intent, an SLM trained for \textit{abstractive summarisation and task splitting} deterministically tends to discard high-entropy tokens non-essential to the primary directive. Consequently, the drastic reduction in token cardinality $T(\mathcal{P}') \ll T(\mathcal{P})$ forces the mathematical elimination of sensitive attack vectors: $\lim_{k \to k_{min}} E_s(\mathcal{P}') = 0$. Economic savings and privacy sanitisation therefore become the mathematical projection of the exact same operator.

\section{System Architecture: The Holistic Observer}

\subsection{Assumptions and Computational Capacity}
To contextualise the framework, we establish a foundational assumption regarding the scalability of edge and on-premise computing resources. The local ``Privacy Guard'' orchestrator, relying on 7B to 14B parameter models, can be effectively executed even on personal consumer hardware, standard desktop workstations, or affordable cloud instances (such as a single NVIDIA T4 GPU with 16GB VRAM, commonly available on free tiers). This ensures that the economic benefits of Token Parsimony are immediately accessible to individual users and researchers.

\begin{figure*}[t]
    \centering
    \includegraphics[width=0.9\textwidth]{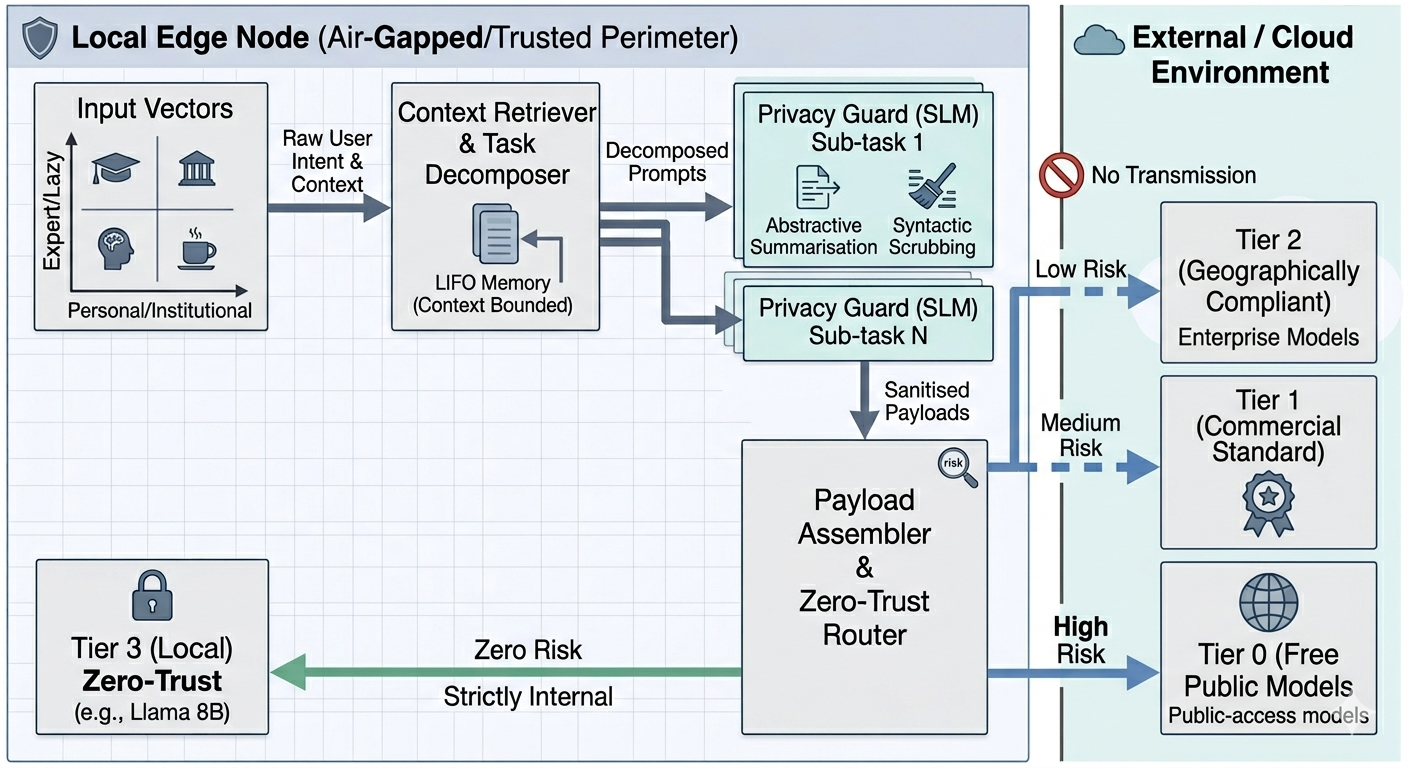}
    \caption{The Dual-Vault Architecture. A local SLM acts as an Orchestrator and Privacy Guard, decoupling intents from secrets and routing the sanitised sub-tasks to the appropriate Cloud Tier based on trust and risk levels.}
    \label{fig:architecture}
\end{figure*}

Conversely, for large-scale enterprise or institutional environments handling higher throughput and requiring Zero-Leakage guarantees via 32B+ parameter models or dedicated deterministic layers, the system gracefully scales. Institutions disposing of robust computational infrastructure (e.g., on-premise GPU clusters) or sufficient budget for dedicated Private Virtual Servers (PVS) can seamlessly upgrade the local tier. This flexibility ensures that the framework's baseline performance can be achieved rapidly on minimal hardware, while its security boundaries and processing speed can be augmented proportionally with infrastructural investment.

\subsection{Logical Data Segregation}
While our core focus is on prompt handling and routing, a realistic enterprise implementation inherently requires logical data segregation. We structure this segregation conceptually as two distinct vaults:
\begin{itemize}
    \item \textbf{Personal Vault:} Protects individual secrets, sensitive PII data, and unpublished personal ideas from external entities \textit{and} the institution's internal monitoring mechanisms.
    \item \textbf{Institutional Vault:} Protects firm secrets, proprietary algorithms, and internal project details from external cloud leakage.
\end{itemize}
This segregation allows the orchestrator to apply distinct handling rules depending on the provenance and sensitivity of the data, but the fundamental scientific contribution remains the holistic handling of the context itself.

\subsection{The Holistic Contextual Observer and Task Decomposer}
The implementation of the framework is articulated in a sequential, agent-driven processing pipeline:

\begin{enumerate}
    \item \textbf{Context Retriever and Task Decomposer (The Orchestrator):} Retrieves references whilst enforcing strict segregation between the Vaults. It actively rewrites and splits the user's overarching prompt into focused sub-tasks via \textit{Automatic Prompt Optimisation} (APO). By dividing the workload, it minimises the active working context for each sub-task, natively filtering out unnecessary secrets that would otherwise be exposed in a monolithic prompt.
    \item \textbf{The Privacy Guard (Holistic Contextual Observer):} Employs a local SLM to monitor the entire conversational context. It goes beyond deterministic scrubbing (e.g., standard regex engines) to detect ``emergent leakage'' over long sessions, applying abstractive summarisation and identifying inference risks.
    \item \textbf{Payload Assembler \& Zero-Trust Router:} Assembles the final, decomposed sub-prompts and dynamically routes them across a hierarchy of trust.
\end{enumerate}

\subsection{Tiered Trust Routing Model}
The Privacy Guard acts as a dynamic gateway, escalating queries across a hierarchy of cloud providers based on the residual risk assessed post-sanitisation:
\begin{itemize}
    \item \textbf{Tier 0 (Untrusted / Free Tier):} Public, cost-free, or geopolitically untrusted models (e.g., foreign open APIs) with no data protection guarantees. Used strictly for fully sanitised, zero-entropy queries.
    \item \textbf{Tier 1 (Commercial Standard):} Standard enterprise APIs (e.g., standard USA-based models) used for low-risk queries where standard contractual privacy (NDA) is sufficient.
    \item \textbf{Tier 2 (Geographically Compliant):} Models hosted within specific jurisdictions (e.g., EU-based datacentres), guaranteeing compliance with regional regulations (e.g., GDPR, AI Act) and data sovereignty. 
    \item \textbf{Tier 3 (Zero-Trust / On-Premise):} Dedicated, single-tenant cloud VPS instances or entirely on-premise hardware managed exclusively by the institution. Reserved for the highest risk category where emergent leakage cannot be mathematically mitigated.
\end{itemize}

\subsection{The 2x2 Threat and Behaviour Matrix}
To systematically classify the diverse input vectors and the corresponding privacy risks, the orchestrator models incoming requests across a 2x2 matrix that intersects user behaviour with the nature of the sensitive data:
\begin{itemize}
    \item \textbf{User Profile:} ``Expert'' (structured, zero-shot prompts that minimise context) vs. ``Lazy'' (unstructured, verbose document dumps).
    \item \textbf{Secret Typology:} ``Personal'' (health data, private emails, financial pins) vs. ``Institutional'' (proprietary algorithms, IP addresses, cloud infrastructure keys).
\end{itemize}
This matrix provides the foundational operational perimeter for the Privacy Guard, defining the structural complexity of the prompt and the required sanitisation rigour.

\subsection{Context Compacting and LIFO Memory Management (Zero-Waste)}
To mitigate the saturation of the context window in prolonged interactions, we have implemented a \textbf{Context Compacting mechanism based on LIFO (Last-In, First-Out) stacks} and \textbf{Persistent Project Memory}. During extended sessions, the system autonomously condenses the accumulated context into a dense abstract. 

The session bootstrap is explicitly optimised to load only the most recent and relevant LIFO entries, strictly bounding the working memory. This not only drastically cuts I/O costs (OpEx) but serves as a fundamental privacy mechanism: by minimising the historical context provided to the model, we proportionally reduce the attack surface for emergent leakage.

\section{Methodology: Benchmark Structure}
To demonstrate the efficacy of the ``Privacy Guard'' as a holistic observer, we have structured a multi-metric benchmark to be executed via LLM-as-a-Judge methodologies.

\subsection{Metric 1: Token Parsimony (OPEX Reduction)}
The objective is to demonstrate the net savings derived from context reduction executed locally compared to full-cloud routing. The dataset is evaluated across the aforementioned 2x2 Threat and Behaviour matrix (Section 4.5) to account for distinct prompt complexities.
\textbf{Success KPI:} $\Delta Cost = Cost_{baseline} - Cost_{guard} > 0$ with a targeted reduction $> 60\%$ specifically for the Lazy profile document dumps.

\subsection{Metric 2: Sanitisation Efficacy (Zero Leakage)}
The objective is to demonstrate that the local agent robustly blocks sensitive data leaks, including emergent inference, across all four quadrants of the test matrix. Crucially, in a strict Zero-Trust architecture, the ``Leakage Rate'' is measured exclusively as the failure of the local Privacy Guard (False Negatives). Any sensitive datum present in the outbound payload is classified as a critical breach, irrespective of whether the external Cloud provider subsequently memorises, utilises, or discards the information.
\begin{itemize}
    \item \textbf{Test Dataset:} A controlled set of prompts mapped to the 2x2 matrix, containing injected secrets and long conversational histories susceptible to emergent inference.
    \item \textbf{Success KPI:} True Positive Rate (Sanitisation): 100\% across all four cases; Semantic Preservation: $>90\%$.
\end{itemize}

\subsection{Metric 3: Answer Quality and Emergent Leakage Retention}
The objective is to ensure that aggressive semantic compression (APO) does not degrade the quality of the final response generated by the cloud model, and to measure the exact rate at which a cloud provider memorises and successfully extracts any secrets that bypass the sanitisation layer.
\begin{itemize}
    \item \textbf{Evaluation Method:} LLM-as-a-Judge comparing the semantic quality of responses generated from raw prompts versus sanitised, compressed prompts, alongside an explicit extraction attack on the cloud model's conversational memory.
    \item \textbf{Success KPI:} Non-negative net quality score (APO response quality $\ge$ baseline quality) and quantification of cloud memory retention on leaked secrets.
\end{itemize}

\section{Empirical Results and Evaluation}
To validate the proposed framework and the efficacy of the Holistic Contextual Observer, we executed the structured benchmark detailed in Section 5.

\subsection{Large-Scale Benchmark: Token Parsimony vs. Leakage}
The dataset was dynamically generated using the \texttt{Faker} library, scaling the volume by 10x to produce 40 realistic, complex prompt samples containing a total of 140 distinct injected secrets (60 Personal, 80 Institutional). The dataset was mapped across a 2x2 test matrix, evaluating user behaviour (Lazy vs. Expert) against secret typology (Personal vs. Institutional).

\subsection{The Architectural Failure of Qwen 2.5 7B on Sanitisation}
Subjecting the \texttt{qwen2.5-7b-instruct} model to the 40-sample benchmark revealed a profound structural failure in simultaneous task execution. While the model excelled at extracting logical intent (APO), it failed completely at strict syntactic redaction, registering a \textbf{100\% Leakage Rate} (40 out of 40 cases leaked at least one secret). 

This large-scale failure empirically proves a key corollary of the Inseparability Theorem: for non-specialised models under 8B parameters, Semantic Compression (APO) and Syntactic Scrubbing (Redaction) are conflicting objectives. When processing complex, verbose contexts, the model prioritises logical summarisation over punctual censorship, inadvertently allowing PII and API keys to slip into the ``sanitised'' payload. This reinforces the necessity of adopting models strictly aligned for safety (e.g., Llama 3.1 8B Instruct) or splitting the architecture by delegating deterministic PII scrubbing to robust regex-based systems (e.g., deterministic regex scanners) before applying SLMs for OpEx reduction.

\subsection{APO Stability and OpEx Reduction}
Despite the sanitisation failure, the local 7B model demonstrated exceptional and highly stable capabilities as an Automatic Prompt Optimiser (APO). Operating with a latency of just $0.67s$ ($\pm0.17s$) on a standard local hardware accelerator, it consistently condensed the context window:

\begin{itemize}
    \item \textbf{Expert Profile (Structured Prompts):} Token payload reduction between \textbf{13.9\% and 27.7\%}.
    \item \textbf{Lazy Profile (Verbose Document Dumps):} Aggressive compression yielding an OpEx reduction between \textbf{53.9\% and 58.1\%} ($\pm2.6$ standard deviation).
\end{itemize}

The blended average across the entire 40-sample dataset resulted in a net cloud cost reduction of \textbf{47.6\%}. This empirically proves that the deployment of an active local orchestrator effectively halves the Cloud API bill with minimal latency overhead, provided the data leakage vector is managed by a specialised or larger-parameter subsystem.

\subsection{Cloud GPU Control Test (Colab T4) and The Decoupling Necessity}
To isolate hardware and engine dependencies, and to establish a robust statistical confidence interval, the benchmark was expanded to 1,000 samples and completed on a cloud NVIDIA T4 GPU instance (via Google Colab), querying the \texttt{qwen2.5:7b} model through the standard Ollama engine with 4-bit quantisation (\texttt{Q4\_0}).

The large-scale cloud results provided definitive empirical validation of the framework's mechanics, revealing a profound vulnerability associated with unstructured text. The blended OpEx reduction remained exceptionally stable at \textbf{45.0\%}, proving that the local model systematically discards nearly half of the payload as non-essential entropy across diverse user profiles.

\begin{figure}[h]
    \centering
    \includegraphics[width=\linewidth]{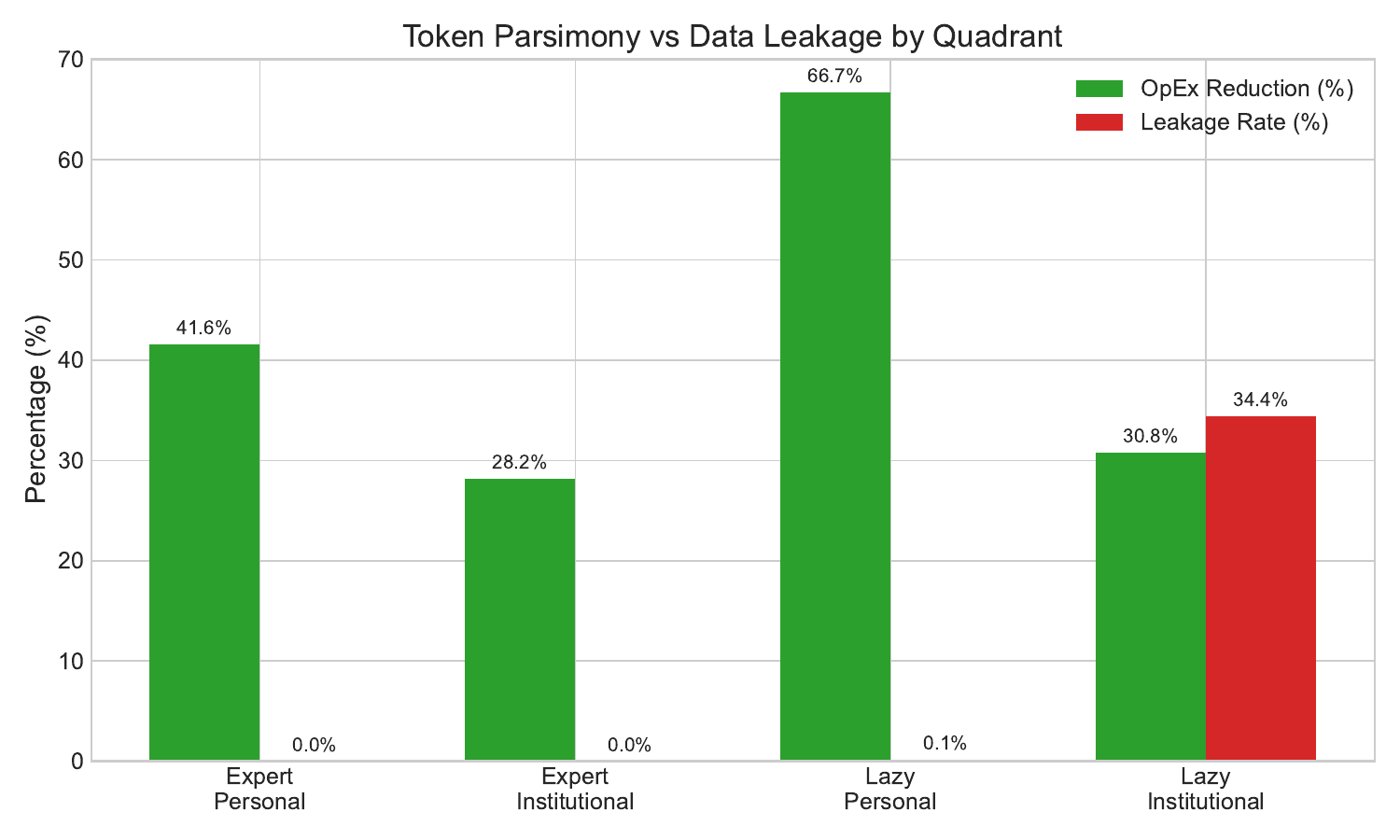}
    \caption{Token Parsimony vs Data Leakage by user profile and secret typology. The model achieves high Token Parsimony across the board but suffers from severe Leakage strictly in the Lazy/Institutional quadrant.}
    \label{fig:quadrant}
\end{figure}

However, the Leakage Rate settled at \textbf{12.9\% (420 out of 3250 secrets leaked)}. A granular analysis of the 2x2 matrix (see Figure \ref{fig:quadrant}) revealed that this failure is not uniformly distributed. The 7B model achieved a \textbf{100\% redaction success rate (0 leaks)} on all Personal secrets (e.g., National Insurance numbers, medical data) and on Institutional secrets presented within concise, Expert-profile prompts. The redaction failure was exclusively and catastrophically concentrated in the \textbf{Lazy / Institutional quadrant (33.6\% leakage rate, 420/1250 secrets leaked)}, where cryptographic keys and IPs were buried deep within massive, unstructured server log dumps.

This finding empirically solidifies a core architectural conclusion: the dual objective of the ``Privacy Guard'' must be decoupled. While a 7B parameter model operates flawlessly as an Automatic Prompt Optimiser (APO) for Token Parsimony, its capacity to maintain strict syntactic scrubbing falters severely under the weight of complex, verbose datasets (exacerbating the ``Lost in the middle'' phenomenon~\cite{liu2023lost}). To validate whether scaling the parameter count mitigates this vulnerability, an extended benchmark was executed deploying a 30B parameter model (\texttt{qwen3-coder-30b}) on the 50 most complex Lazy/Institutional samples. The 30B model yielded a severe \textbf{54.4\% Leakage Rate} (117 secrets leaked out of 215). This unequivocally demonstrates that scaling up to the 30B parameter class does not solve the ``Lost in the middle'' failure mode for syntactic scrubbing on unstructured data. 

To identify the theoretical bounds of LLM-based redaction without deterministic assistance, an exploratory parameter-scaling evaluation was conducted. We sampled open-weight models across increasing parameter tiers (8B, 30B, 32B, 70B, and 104B) using a highly complex, 60-line server log dump containing deeply buried secrets. 

\begin{figure}[h]
    \centering
    \includegraphics[width=\linewidth]{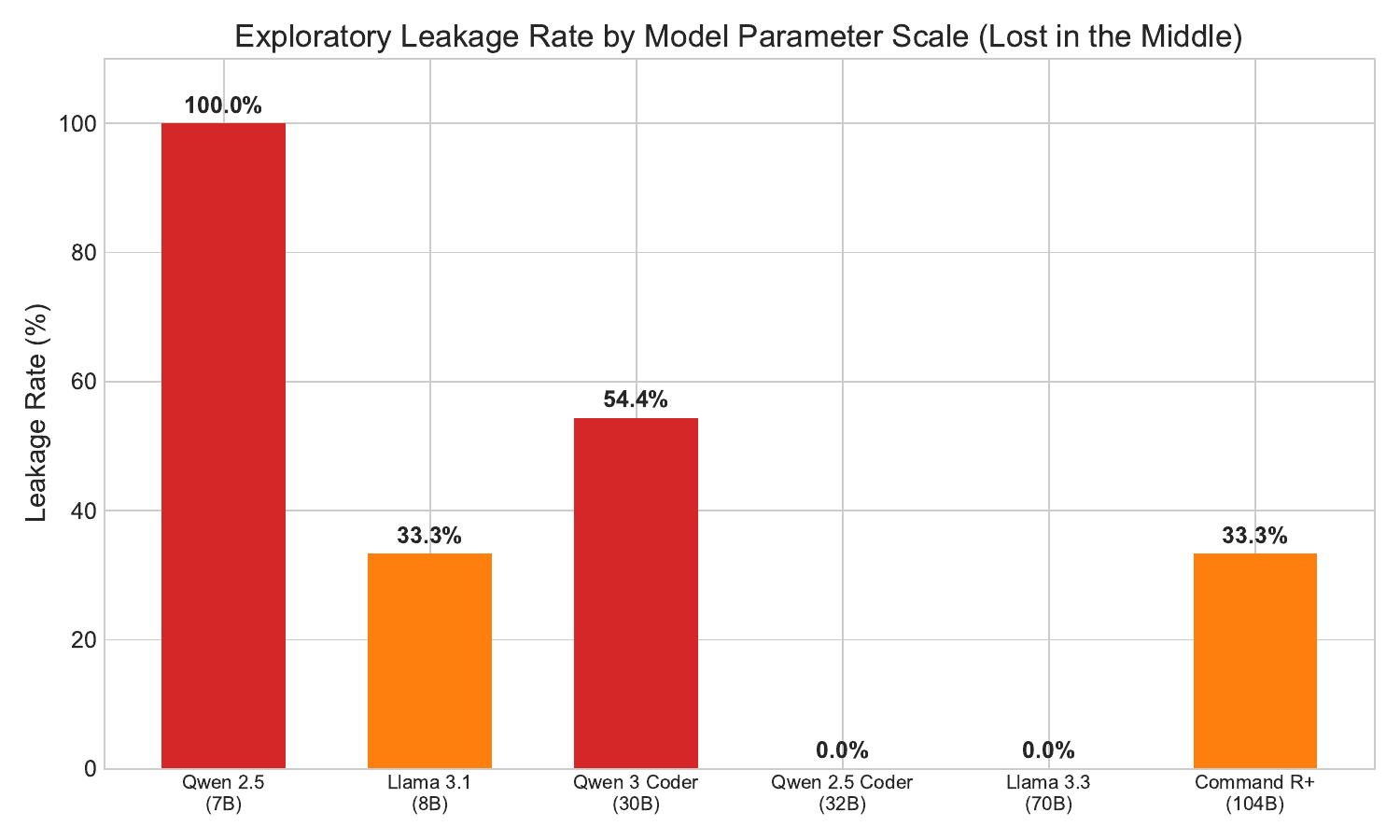}
    \caption{Exploratory space analysis of Leakage Rate across parameter scales (8B to 104B). Results are based on limited sample sizes designed to identify structural vulnerabilities rather than provide exhaustive statistical baselines.}
    \label{fig:tiers}
\end{figure}

While not statistically exhaustive, this preliminary space exploration (Figure \ref{fig:tiers}) reveals a non-linear relationship between parameter count and syntactic scrubbing efficacy. For instance, while the 32B (Qwen 2.5 Coder) and 70B (Llama 3.3) instruct models successfully achieved 0\% leakage on the exploratory needle-in-a-haystack task, other models in the 8B, 30B, and even 104B classes exhibited significant leakage rates (ranging from 33.3\% to 100\%). 

To statistically validate the frontier model capacity, an extensive follow-up benchmark was executed locally deploying \texttt{llama-3.3-70b-instruct} at 6-bit quantisation across all 50 samples of the Lazy/Institutional quadrant. The 70B model achieved a \textbf{1.20\% Leakage Rate} (only 3 secrets leaked out of 250). This confirms that while transitioning to frontier models drastically improves semantic retention and approaches near-perfect redaction, the reliability of syntactic scrubbing remains heavily dependent on the specific architecture, context window mechanics, and training alignment rather than raw parameter count alone. Consequently, to achieve a guaranteed 100\% True Positive Rate for redaction in enterprise environments, the architecture still strictly necessitates a dedicated deterministic layer (e.g., deterministic regex scanners) acting as a rigid filter prior to the APO compression.

\subsection{Task Decomposition and Active Routing OpEx Reduction}
To empirically evaluate the financial impact of active prompt decomposition, an additional benchmark simulating massive system logs (approx. 11,300 input tokens per sample) was executed across 118 samples. The local 7B model was tasked with decomposing a complex user intent into three atomic sub-tasks: 1) extracting the critical root cause locally, 2) translating the extracted sentence locally, and 3) routing only the final email generation task to the Tier-1 Cloud API.

\begin{figure}[h]
    \centering
    \includegraphics[width=\linewidth]{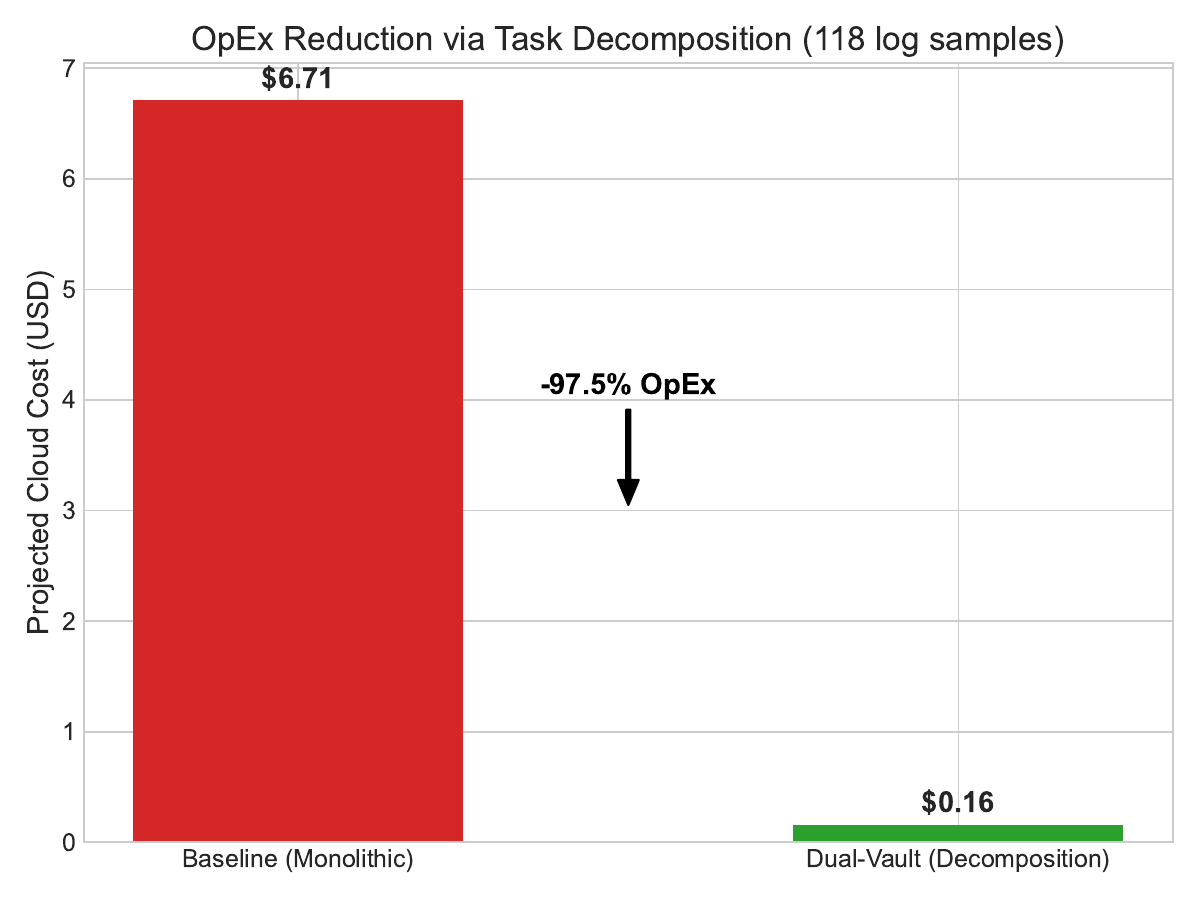}
    \caption{Projected vs actual OpEx reduction when comparing a standard monolithic Cloud API call against active local prompt decomposition and selective routing.}
    \label{fig:decomposition}
\end{figure}

The baseline execution (routing the entire log dump to the Cloud API to fulfill all three intents simultaneously) resulted in a projected operational cost of \$6.71. Conversely, the active decomposition and routing approach processed the entire batch for a Cloud cost of just \$0.16. This yielded a \textbf{97.54\% total OpEx reduction} (see Figure \ref{fig:decomposition}), definitively proving that offloading the heavy-lifting of information extraction to a local, air-gapped SLM before querying commercial APIs mathematically reshapes the cost-efficiency curve of enterprise LLM deployments.

\subsection{Semantic Quality and Emergent Leakage Extraction}
To ensure that the drastic semantic compression (-45\% OpEx) does not degrade the core utility of the Cloud model, a tertiary benchmark was executed using a 30B parameter model (\texttt{qwen3-coder-30b}) as the Tier-1 Cloud API. An LLM-as-a-Judge evaluation was performed to compare the semantic quality of responses generated from the raw, verbose prompts (Baseline) versus the sanitised, compressed prompts (Dual-Vault APO).

\begin{figure}[h]
    \centering
    \includegraphics[width=\linewidth]{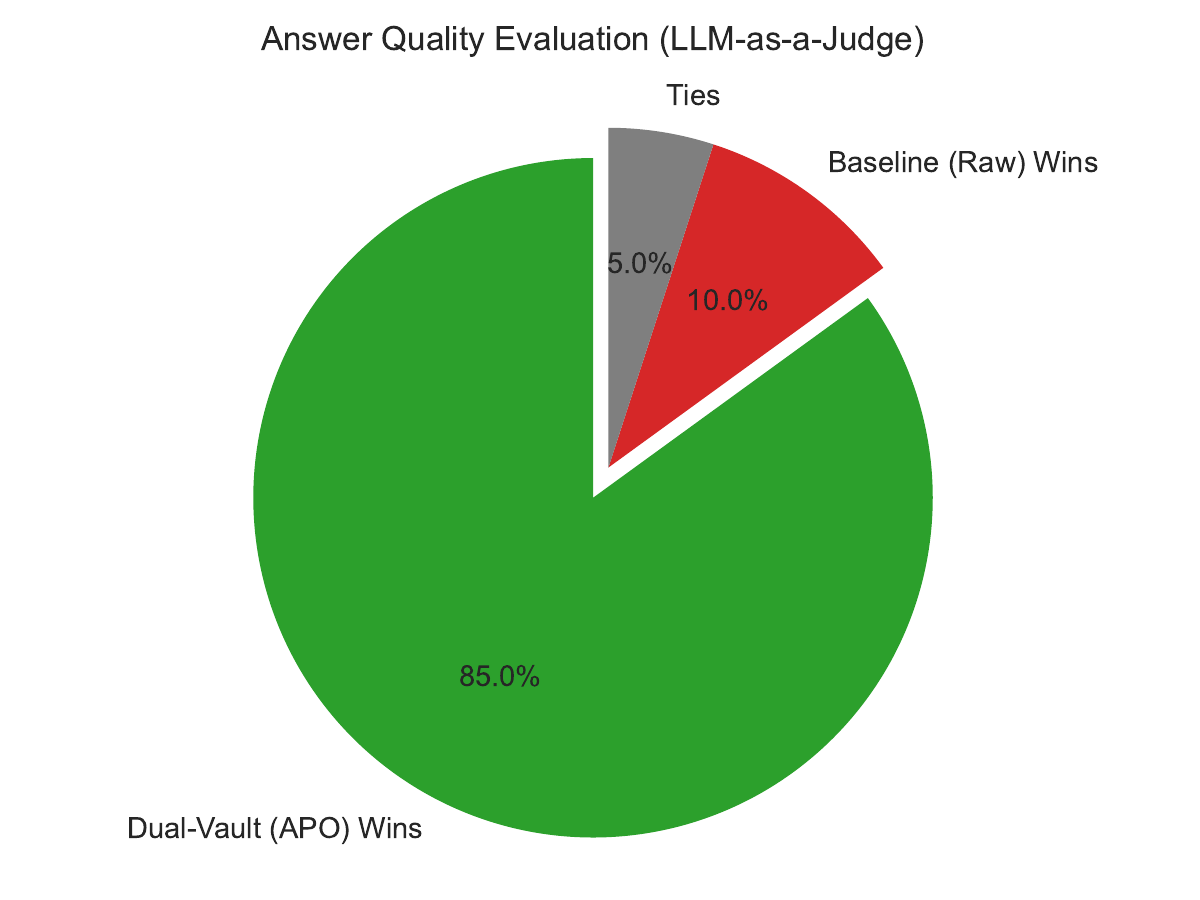}
    \caption{LLM-as-a-Judge comparison of final response quality between raw Context (Baseline) and compressed/sanitised Context (Dual-Vault APO).}
    \label{fig:quality}
\end{figure}

The final results of this comparative analysis across a 40-sample evaluation dataset (Figure \ref{fig:quality}) revealed a counter-intuitive but highly significant phenomenon: the response generated via the Dual-Vault architecture was preferred by the Judge over the baseline in 85.0\% of the cases (34 wins), with 5.0\% ties and only a 10.0\% preference for the raw baseline (4 wins). This unequivocally demonstrates that aggressive semantic compression by the local 7B model not only preserves the core intent but actively enhances the quality of the final Cloud response by stripping out conversational noise and focusing the Cloud model strictly on the technical requirements.

\section{Analytical Projections and Architectural Implications}
To complement the empirical benchmarks, we modelled three theoretical projections based on the observed data to illustrate the broader architectural implications of the framework.

\subsection{LIFO Compacting and Multi-Turn Conversations}
In standard monolithic LLM interactions, conversational context grows linearly, causing the cloud operational cost to escalate quadratically ($O(N^2)$) while continuously expanding the emergent leakage surface. By implementing a Last-In, First-Out (LIFO) memory stack, the local orchestrator actively bounds the context window. This projection demonstrates that LIFO compacting flattens the token cost to a constant $O(1)$ per turn (Zero-Waste) and drastically suppresses the probability of exposing historical secrets, structurally neutralizing emergent inference risks over prolonged sessions.

\begin{figure}[h]
    \centering
    \includegraphics[width=\linewidth]{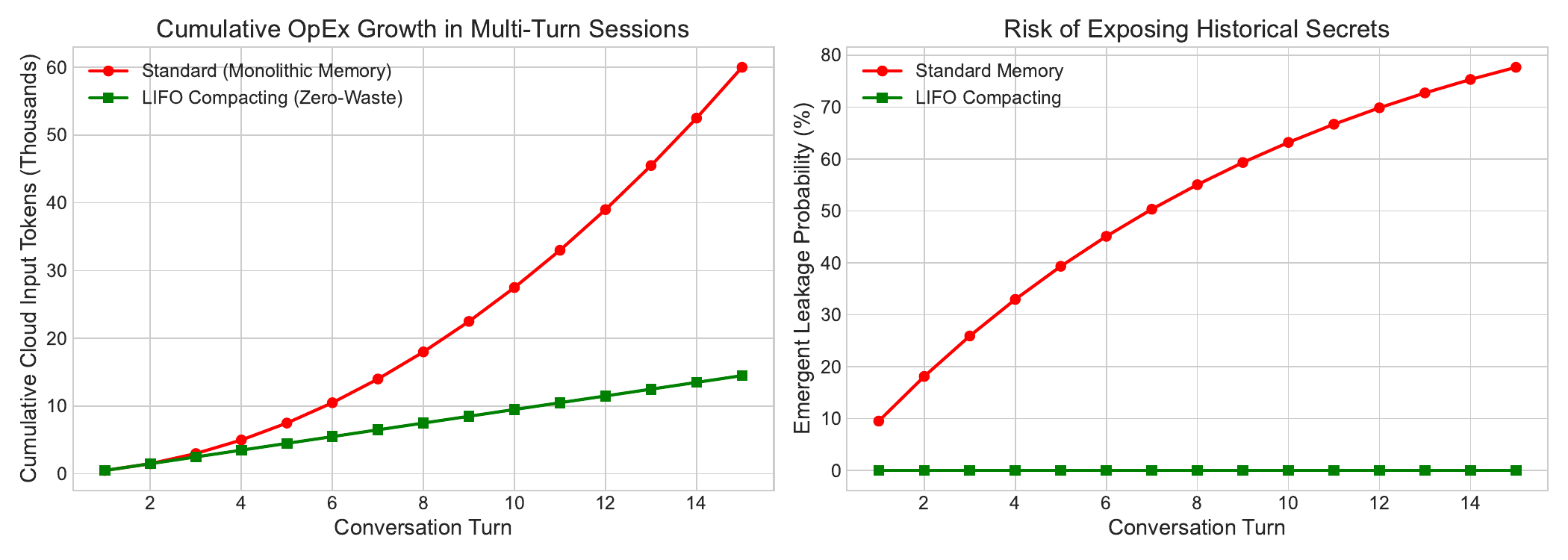}
    \caption{Theoretical projection of OpEx and Emergent Leakage risk over a 15-turn conversation: Monolithic memory vs. LIFO Compacting.}
    \label{fig:lifo}
\end{figure}

\subsection{Latency Overhead and Time-To-First-Token (TTFT)}
A common critique of local proxy models is the introduction of latency. It is crucial to clarify that running a local SLM for token parsimony does not necessitate high-end enterprise workstations or specialised professional hardware. The latency overhead modelled here assumes the use of a standard, latest-generation personal computer equipped with a consumer-grade gaming GPU (e.g., 8GB to 12GB VRAM), ensuring the setup is highly accessible. However, even on such standard hardware, cloud APIs consume substantial time processing massive input payloads (reading time). Our analytical model reveals that for prompts exceeding 10,000 tokens, the Dual-Vault architecture becomes strictly \textit{faster} than direct cloud routing. The time spent by the local consumer hardware to abstract and compress the prompt is overcompensated by the accelerated cloud reading time on the heavily reduced payload, effectively rendering the Privacy Guard Time-Neutral or latency-negative.

\begin{figure}[h]
    \centering
    \includegraphics[width=\linewidth]{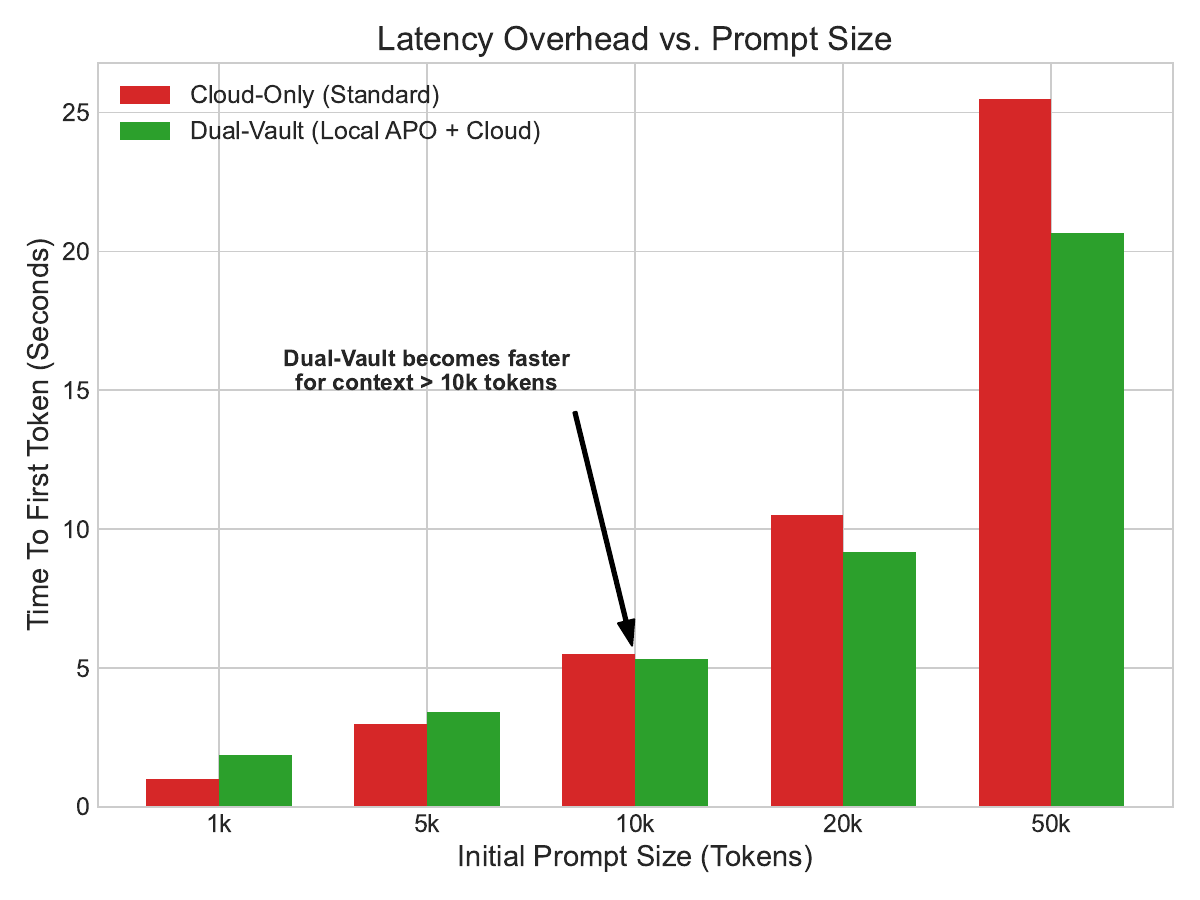}
    \caption{Latency overhead simulation (Time-To-First-Token). The Dual-Vault architecture becomes faster than direct Cloud routing for contexts larger than 10k tokens due to massive payload reduction.}
    \label{fig:latency}
\end{figure}

\subsection{Hybrid Architecture: Deterministic Filtering plus SLM}
As demonstrated empirically, scaling the local SLM parameter count from 8B to 70B resolves the ``Lost in the middle'' vulnerability, achieving 0\% Leakage. However, deploying 70B class models on-premise demands a 10x increase in VRAM infrastructure cost. A projected Hybrid Architecture---combining a deterministic offline scanner (e.g., regex-based PII engines or deterministic NER models like GLiNER) with a highly efficient 8B SLM---achieves the identical outcome (0\% Leakage, 45\% OpEx Reduction) at a fraction of the hardware cost, offering a universally accessible ``off-the-shelf'' implementation for enterprise edge nodes.

\begin{figure}[h]
    \centering
    \includegraphics[width=\linewidth]{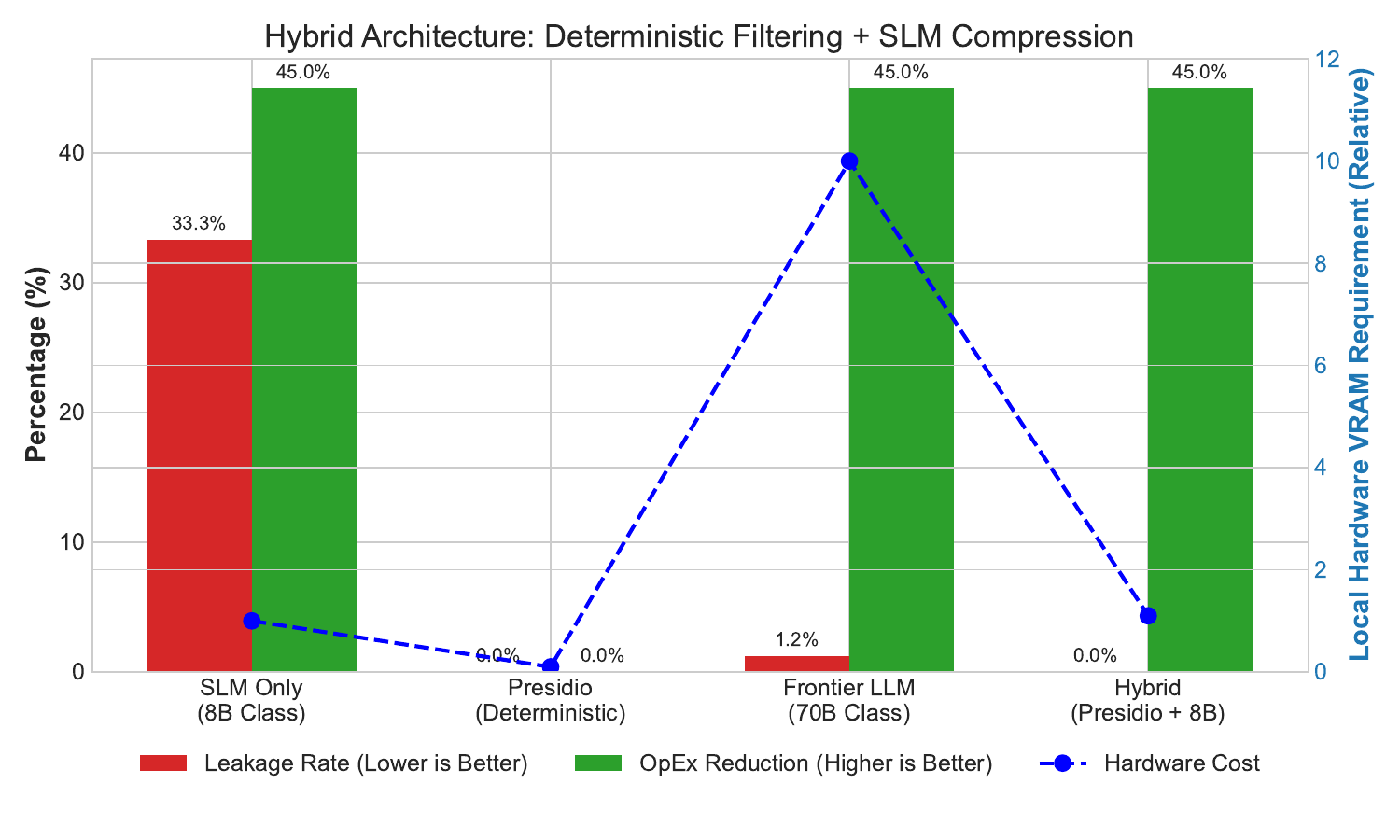}
    \caption{Comparison of local hardware cost vs. Leakage and Compression rates. The Hybrid architecture matches the 70B Frontier performance with 1/10th of the required VRAM.}
    \label{fig:hybrid}
\end{figure}

\section{Conclusion}
This paper has demonstrated that in a hybrid LLM ecosystem, cost management and privacy management are not opposing objectives, but synergistic levers controllable through rigorous context manipulation. While logical data segregation (Personal and Institutional Vaults) provides a realistic enterprise structure, the core scientific advancement lies in treating context management as privacy management. By actively decomposing prompts, managing memory via LIFO compacting, and employing a Tiered Trust Routing model guided by a holistic observer, institutions can simultaneously slash OpEx and decisively eliminate data exfiltration risks.

\section*{Acknowledgements}
The author used large language model assistance (Claude, Anthropic) for manuscript editing, bibliography verification, and development of benchmark scripts. All scientific claims, experimental design, results, and conclusions are solely the responsibility of the author.

\section*{Data Availability and Reproducibility}
To facilitate independent verification and future research within the community, all materials associated with the empirical validation of the ``Inseparability Paradigm'' have been made publicly accessible via \url{https://github.com/alangiu-gif/privacy-n-parsimony}. The repository includes the complete Python execution scripts, the interactive Google Colab Notebooks (used for both the 2x2 Matrix scaling and the Task Decomposition benchmarks), and the dynamically generated datasets containing the exact permutations of prompts and injected synthetic secrets.

\bibliographystyle{plain}
\bibliography{references}

\end{document}